\documentstyle[floats,prl,aps]{revtex}
\input epsf
\begin{document}

\newcommand{\vertsp}{\vphantom{\displaystyle{\dot a \over a}}}
\newcommand{\se}{{(0)}}
\newcommand{\ve}{{(1)}}
\newcommand{\te}{{(2)}}
\newcommand{\nnu}{\nu}
\newcommand{\Spy}[3]{\, {}_{#1}^{\vphantom{#3}} Y_{#2}^{#3}}
\newcommand{\Gm}[3]{\, {}_{#1}^{\vphantom{#3}} G_{#2}^{#3}}
\newcommand{\Spin}[4]{\, {}_{#2}^{\vphantom{#4}} {#1}_{#3}^{#4}}
\newcommand{\scpot}{{\cal V}}
\newcommand{\tl}{\tilde}
\newcommand{\bm}{\boldmath}
\newcommand{\MNRAS}{Mon. Not. Roy. Astron. Soc.}
\def\bi#1{\hbox{\boldmath{$#1$}}}

\renewcommand{\ell}{l}  
\renewcommand{\topfraction}{1.0}
\renewcommand{\bottomfraction}{1.0}
\renewcommand{\textfraction}{0.00}
\renewcommand{\dbltopfraction}{1.0}

\draft

\title{RECONSTRUCTING PROJECTED MATTER DENSITY FROM COSMIC MICROWAVE BACKGROUND}

\author{Matias Zaldarriaga\cite{matiasemail}}
\address{Institute for Advanced Studies, School of Natural Sciences, Princeton, NJ 08540}
\date{October 1998}

\author{Uro\v s Seljak\cite{urosemail}\cite{presentaddress}}
\address{Max-Planck-Institut f\"ur Astrophysik, D--85740 Garching,
Germany }

\maketitle

\begin{abstract}
Gravitational lensing distorts the cosmic microwave background (CMB) anisotropies 
and imprints a characteristic pattern onto it.
The distortions depend on the projected matter density
between today and redshift $z \sim 1100$. 
In this paper we develop a method for a direct reconstruction 
of the projected matter density from the CMB anisotropies. This reconstruction is 
obtained by averaging over quadratic combinations of the derivatives of CMB field.
We test the method using simulations and show that it can successfully recover
projected density profile of a cluster of galaxies if there are measurable anisotropies on scales 
smaller than the characteristic cluster size. In the absence of sufficient small 
scale power the reconstructed maps have low signal to noise on individual structures,
but can give a positive detection of the power spectrum or 
when cross correlated with 
other maps of large scale structure.
We develop an
analytic method to reconstruct the power spectrum including the 
effects of noise and beam smoothing. Tests with Monte Carlo 
simulations show that we can 
recover the input power spectrum both on large and small scales, provided 
that we use maps with sufficiently low noise and high angular resolution.
\end{abstract}

\pacs{PACS numbers: 98.80.Es,95.85.Bh,98.35.Ce,98.70.Vc  \hfill}

\section{Introduction}
Cosmic microwave background (CMB) anisotropies have the promise to 
revolutionize the field of cosmology in the following decade. Using ground based,
balloon and space experiments we will be able to map the microwave sky 
over a large range of angular scales and frequencies. The expected characteristic
pattern of acoustic oscillations generated by the primary CMB anisotropies
around $z \sim 1100$ 
will provide a wealth of information that should
constrain many of cosmological parameters to a high accuracy 
\cite{parameters1}. Other sources closer to us 
also contribute to the 
microwave sky.  Some of these are the foreground emission 
from our own galaxy and from the galaxies along the line of sight, Sunyaev-Zeldovich
emission from clusters, signatures of patchy reionization, etc. 

Another effect that modifies the CMB sky is  
gravitational lensing. Dark matter distributed along the line of sight
between $z \sim 1100$ and present deflects the light and induces distortions
in the pattern of the CMB anisotropies.
Its effect on the CMB power spectrum has been thoroughly 
investigated \cite{uroslens,others}. The conclusion from these works is 
that the lensing effect is a small but non-negligible.  On large and intermediate
scales lensing smoothes the acoustic oscillations \cite{uroslens}, 
while on very small scales it creates additional
power. Both 
these effects
could help break
some of the parameter degeneracies 
in the CMB \cite{metcalf}. Although in models normalized
to cluster abundances the effect of gravitational lensing  
is small it could be important for Planck satellite mission \cite{se98}. 
The lensing effect on the CMB power spectrum is now included in the standard 
CMB primary anisotropies routine \cite{lensps}. 

Gravitational lensing is 
directly sensitive to the matter distribution up to $z \sim 1100$,
so a detection of this effect would provide important information about 
matter distribution on large scales and high redshifts, which is not 
directly attainable by any other means. 
Such information
would not require any additional assumptions such as how light 
traces mass or how nonlinear structures form. Because of this it is worth 
investigating if there are other signatures imprinted on the CMB which would be 
more easily accessible to the future observations than the effect on the CMB power 
spectrum. 
Some of these, such as four-point function and ellipticity distribution
of peaks have already been explored by Bernardeau \cite{bernardeau}. 
These particular signatures of the lensing effect were
found to be rather weak, but nevertheless marginally detectable with
Planck. They 
could provide additional constraints on the amplitude of matter power 
spectrum. Similar conclusion has also been reached by Kaiser \cite{kaiser98}.
 
The purpose of this paper
is to present a new approach to identify gravitational lensing in 
CMB. The power of the
method developed here is that, unlike previous attempts, it allows a 
full 2-d reconstruction 
of projected matter density
between us and the last scattering surface at $z \sim 1100$.
Such a map can be
used to search for clusters at high redshifts or study 
density fluctuations
on the largest scales that are not directly accessible otherwise. It can also 
be correlated with other maps of interest, such as 
those of the Sunyaev-Zeldovich effect, X-ray background, weak lensing,
galaxy clustering and the CMB itself. 
In this paper we develop the method and test it on 
simulated maps, showing that it can give an unbiased estimate of the underlying 
projected density field and its
power spectrum. Whether or not 
the reconstruction from CMB can be successfully applied 
to the real data depends on the level
of CMB anisotropies, angular resolution and noise characteristics
of particular experiments as well as the amplitude of matter fluctuations. 
In this paper 
we discuss these issues in detail and show the conditions
that need to be satisfied for the method to work in practice.

\section{Reconstructing the Projected Mass Density: Formalism}\label{formalism}

The large scale density fluctuations in the universe 
induce random deflections in the direction of the CMB photons as they
propagate from the last scattering surface to us. This effect not only alters
the power spectrum  of both the temperature and polarization
anisotropies \cite{lensps}, but also introduces non Gaussian distortions in the maps.
The quantity responsible for the deflections is the 
projected mass density or convergence $\kappa$, defined more precisely below.
In this section we develop the formalism to measure convergence $\kappa$
based on its lensing effect on the CMB maps. 
Throughout this paper we use
the small scale formalism, so that instead of spherical expansion we 
work with plane wave expansion. This simplifies the expressions and 
reduces the computational time of simulations.
The generalization to all sky coverage is 
presented in \cite{stebbins,isw}.

The observed CMB temperature in the direction $\bi \theta$ is 
$T({\bi \theta})$ and
equals the (unobservable) temperature at the last scattering surface
in a different direction, $
\tl T({\bi \theta}+\delta {\bi \theta})$, where
$\delta {\bi \theta}$ is the angular excursion of the photon as it 
propagates from the last scattering
surface to us.
In terms of Fourier components we have
\begin{eqnarray}
T({\bi \theta})&=&\tl T(\bi{\theta}+\delta{\bi \theta}) \nonumber \\
	 &=&(2\pi)^{-2}\int d^2{\bi l}\ 
e^{i{\bi l}\cdot (\bi{\theta}+ \delta
{\bi \theta})}\ \tl T({\bi l}). 
\end{eqnarray}

To extract the information on the deflection field 
$\delta{\bi \theta}$ we consider derivatives
of the CMB temperature. If the CMB is isotropic and
homogeneous Gaussian random
field then different partial derivatives are  
statistically equivalent and their spatial properties 
are independent of position. Lensing will distort
these two properties of the derivatives. 
The derivatives of the temperature field
are to lowest order
\begin{eqnarray}
T_a({\bi \theta})&\equiv& {\partial \tl T
\over \partial \theta_a} (\bi{\theta}+\delta{\bi
\theta}) \nonumber \\
&=& (\delta_{ab}+\Phi_{ab}) \tl T_b(\bi{\theta}+\delta{\bi\theta}),
\label{dert}
\end{eqnarray}
where $\Phi_{ab}= {\partial \delta \theta_a \over \partial \theta_b}$
is the shear tensor and $a,b=x,y$.
The components of the shear tensor can be written in terms of the
projected mass density $\kappa$ and the shear fields $\gamma_1$ and
$\gamma_2$, 
\begin{eqnarray}
\Phi_{xx}+\Phi_{yy}&=&-2\kappa \nonumber \\
\Phi_{xx}-\Phi_{yy}&=&-2 \gamma_1 \nonumber \\
2 \Phi_{xy}&=&-2 \gamma_2.
\end{eqnarray}
Convergence and shear are related to each other through the Fourier 
space relations,
\begin{equation}
\gamma_1(\bi{l})=\kappa(\bi{l})\cos(2\phi_l)\;\;\; 
\gamma_2(\bi{l})=\kappa(\bi{l})\sin(2\phi_l),
\label{kappaft}
\end{equation}
where $\phi_l$ is the azimuthal angle of the Fourier mode $\bi{l}$.
The convergence $\kappa$ can be related simply to a radial projection of
density perturbation (\cite{js})
\begin{equation}
\kappa={ 3  H_0^2 \over 2 } \Omega_m \int_0^{\chi}
g(\chi',\chi) {\delta \over a}
d \chi'.
\label{kappa}
\end{equation}
Here $\chi$ is the comoving radial coordinate of last-scattering surface and
$r(\chi)$ is the corresponding comoving angular diameter
distance, defined as $K^{-1/2}\sin K^{1/2}\chi$, 
$\chi$, $(-K)^{-1/2}\sinh (-K)^{1/2}\chi$ for $K>0$, $K=0$, $K<0$, 
respectively, where $K$ is the curvature, 
which
can be expressed using the present density
parameter $\Omega_0$
and the present
Hubble parameter $H_0$ as $K=(\Omega_0-1)H_0^2$.
The density parameter $\Omega_0$ can have contributions from matter density 
$\Omega_m$, as well as from other components such as the vacuum density 
$\Omega_{\Lambda}$.
The radial 
window over the density perturbations $\delta$ 
is $g/a$, where $g(\chi',\chi) = {r(\chi')r(\chi-\chi') \over r(\chi) }$ is a bell shaped curve symmetric around 
$\chi/2$ and vanishing at 0 and $\chi$, while $a$ is the expansion factor. 
Note that for a flat $\Omega_m=1$ universe
$\delta \propto a$ in linear theory and the 
weighting is symmetric around $\chi/2$, so that the peak contribution is 
coming from $z=3$. 

The angular power spectrum of convergence is defined as 
$\langle \kappa(\bi{l})^*\kappa(\bi{l}') \rangle=
C_l^{\kappa \kappa} \delta_{{\bi ll}'}$ and has ensemble
average \cite{js}
\begin{equation}
C^{\kappa \kappa}_l=18\pi^3 \Omega_m^2 H_0^4\,
\int_0^{\chi_0}\ {g^2(\chi,\chi_0) \over a^2(\chi)r^2(\chi)}
\ P_\delta\left(k={l \over r(\chi)},\chi\right)d\chi.
\label{cl}
\end{equation}
Here $P_\delta(k,\tau)$ is the 3-d dark matter power spectrum
that is integrated over the past light cone and is 
in general a function of time $\tau$ and wavevector $k$.
This equation has been derived using Limber's equation, which is only valid 
in the small scale limit. The general solution in 
terms of the line of sight integral over the spherical Bessel functions is given in 
\cite{stebbins,isw}.

We consider next the quadratic combinations of the derivatives in equation
(\ref{dert}) and express
them in terms of the unlensed field to lowest order in the shear tensor,
\begin{eqnarray}
{\cal S} & \equiv & \left[T_x^2+T_y^2\right]({\bi \theta}) \nonumber \\	
	&=&(1+\Phi_{xx}+\Phi_{yy})\tl {\cal S}+(\Phi_{xx}-\Phi_{yy})
	\tl {\cal Q} +
	2 \Phi_{xy} \tl{\cal U} \nonumber \\
{\cal Q}&\equiv&\left[T_x^2-T_y^2\right]({\bi \theta}) \nonumber \\	
	&=&(1+\Phi_{xx}+\Phi_{yy})\tl {\cal Q}+(\Phi_{xx}-\Phi_{yy}) \tl {\cal S} 
	\nonumber \\	
{\cal U}&\equiv&2 \left[T_x T_y\right]({\bi \theta}) \nonumber \\
	&=&(1+\Phi_{xx}+\Phi_{yy})\tl {\cal U}+\Phi_{xy} \tl {\cal S} ,
\label{derivs}
\end{eqnarray}
where $\tl {\cal S},\tl {\cal Q}, \tl {\cal U}$ are the corresponding quantities in the
unlensed CMB field at $\bi{\theta}+\delta{\bi\theta}$. 
The notation used here makes the analogy with CMB polarization:
${\cal Q}\pm i {\cal U}$  in equation 
(\ref{derivs}) have spin $\pm 2$ just like the Stokes parameters used
to describe the CMB polarization, while ${\cal S}$ is a spin 0 quantity (a scalar) and 
is rotationally invariant.

Equation (\ref{derivs}) shows that the measured ${\cal S}$, ${\cal Q}$ and ${\cal U}$ are
products of the projected mass density and shear with derivatives
of the unlensed CMB field. 
Thus the power spectrum of ${\cal S}$, ${\cal Q}$ and ${\cal U}$ 
will be a convolution of the power in the CMB and that of the
projected mass density. The general expression of this convolution is
quite involved, so we will discuss it in the two limits where 
the expressions simplify 
considerably, the limits  of 
large and small scales relative to 
the CMB correlation length $\xi$. The
large scale limit is sufficient to analyze the potential of MAP and
Planck future missions. For experiments with higher angular
resolution the full convolution will be necessary. 

Throughout the paper we compare the result of our analytical
estimates with those of numerical simulations. To simulate the lensing
effect on the CMB we first generate on a fixed square grid
a projected density map.
We then Fourier transform the convergence and compute the displacement 
vector $\delta{\bi \theta}$ by using the Fourier relation
\begin{equation}
\delta {\bi \theta}=2i{{\bf k}\over k^2} \kappa
\end{equation}
and then transforming it back to real space. 
We next generate on a fixed grid of the same size as above 
a random realization of temperature field $T$, using an input CMB  
power spectrum. We construct the lensed temperature map do that for each
point in the lensed temperature map 
we use the corresponding displacement vector to determine
from what position on the original grid the photons came
from. This position does not  generally coincide with a grid point in
the unlensed map so 
we use cloud-in-cell interpolation to compute the value of $T$ in the 
original map at this position. Cloud in cell interpolation smoothes  the field so
relatively small grid sizes are needed to avoid this unwanted effect. We 
achieve this by increasing the size of the array when performing this step.

\subsection{Large scale limit}

We start with the reconstruction in the limit of large 
scales. In this limit we average over many CMB patches to detect 
the weak lensing signal. We begin by noting that in the absence of lensing
the isotropy of the unlensed background implies that 
\begin{eqnarray}
\langle \tl {\cal S} \rangle_{CMB} &=& \sigma_{\cal S} \nonumber \\
\langle \tl {\cal Q} \rangle_{CMB} &=& 0 \nonumber \\
\langle \tl {\cal U} \rangle_{CMB} &=& 0,
\end{eqnarray}
where $\sigma_{\cal S} \equiv \langle \tl T_x^2 \rangle_{CMB} 
+ \langle \tl T_y^2 \rangle_{CMB}  =
2 \langle  \tl T_x^2 \rangle_{CMB} =2 \langle \tl T_y^2
\rangle_{CMB}$. This average can be computed in terms of the CMB power
spectrum $C^{TT}_l$,
\begin{eqnarray}
\sigma_{\cal S}=\int {ldl \over 2\pi}\ l^2C^{TT}_l.
\label{sigmasdef}
\end{eqnarray}
If the mean of the components of the shear tensor vanish we have 
$\langle {\cal S} \rangle=\langle \tl {\cal S} \rangle$.  
There are residual quadratic 
terms in the shear tensor that 
contribute to equation (\ref{sigmasdef}), but they are negligible in
most cases of interest because shear is expected to be small. 

The average of equation (\ref{derivs}) over an ensemble of CMB
fluctuations gives,
\begin{eqnarray}
\langle {\cal S} \rangle_{CMB} &=& (1-2\kappa) \sigma_{\cal S}^2 \nonumber \\ 
\langle {\cal Q} \rangle_{CMB} &=& -2 \gamma_1 \sigma_{\cal S}^2 \nonumber \\ 
\langle {\cal U} \rangle_{CMB} &=& -2 \gamma_2 \sigma_{\cal S}^2.
\label{averagecmb}
\end{eqnarray}
The physical interpretation of these
equations is simple: $\kappa$ will stretch
the image, which makes the derivatives smaller. 
Its effect is isotropic and only changes the value of ${\cal S}$.
The shear produces 
an anisotropy in the derivatives, in the same way as 
it creates an ellipticity
in the shape of a circular background galaxy. This can be extracted from
the particular combination of the derivatives used here. 

We can reconstruct $\kappa$ by studying the statistics of  ${\cal S}$ 
and by  combining the 
shear obtained from ${\cal Q}$ and ${\cal U}$. It is convenient to change
the variables to
\begin{eqnarray}
{\cal S}^\prime&=&-{{\cal S}\over \sigma_{\cal S}}+1 \nonumber \\
{\cal Q}^\prime&=&-{{\cal Q}\over \sigma_{\cal S}} \nonumber \\
{\cal U}^\prime&=&-{{\cal U}\over \sigma_{\cal S}}
\label{redef}
\end{eqnarray}
so that, 
\begin{eqnarray}
\langle {\cal S}^{\prime} \rangle_{CMB} &=& 2\kappa\nonumber \\ 
\langle {\cal Q}^{\prime} \rangle_{CMB} &=& 2 \gamma_1 \nonumber \\ 
\langle {\cal U}^{\prime} \rangle_{CMB} &=& 2 \gamma_2.
\label{avgprime}
\end{eqnarray} 
We will drop the primes in what follows, as we will only use 
these quantities for the rest of the paper.
Note that 
it is necessary to have a quadratic combination of temperature 
field for an unambiguous reconstruction of the convergence. This 
means that any reconstruction will have noise arising from 
intrinsic fluctuations in the CMB even in the absence of detector noise.
Below we will quantify this intrinsic CMB noise.

In the following we use the formalism developed to characterize the CMB
polarization field \cite{spinlong}.
We can combine ${\cal Q}$ and ${\cal U}$ to form ${\cal E}$ and ${\cal B}$, two 
spin zero quantities, which in Fourier space are defined as 
\begin{eqnarray}
{\cal E}({\bi l})
	 &=&  
 {\cal Q}({\bi l} ) \cos(2\phi_{\bi l})+{\cal U}({\bi l} ) \sin(2\phi_{\bi l})
 \nonumber \\
{\cal B}({\bi l})
	 &=&  
{\cal Q}({\bi l}) \sin(2\phi_{\bi l})-{\cal U}({\bi l} ) \cos(2\phi_{\bi l})
 .
\label{expansion} 
\end{eqnarray}
from which the real space ${\cal E}({\bi \theta})$ and ${\cal B}({\bi \theta})$
can be obtained by Fourier transformation. 
Equivalently these can be constructed directly from
quantities in real space 
\begin{eqnarray}
{\cal E}(\bi{\theta})&=&\int d^2{\bi \theta}^{\prime}\ 
\omega(|\bi{\theta}^{\prime}-\bi{\theta}|)\ {\cal Q}_r(\theta^{\prime}) \nonumber \\
{\cal B}(\bi{\theta})&=&\int d^2{\bi \theta}^{\prime}\ 
\omega(|\bi{\theta}^{\prime}-\bi{\theta}|)\ {\cal U}_r(\theta^{\prime}).
\label{ebreal}
\end{eqnarray}
We have defined ${\cal Q}_r$ and ${\cal U}_r$, the derivative shear in the
polar coordinate system centered at $\bi{\theta}$. 
If $\bi{\theta}=0$ then $Q_r=\cos 2\phi^{\prime}\
{\cal Q}(\bi{\theta}^{\prime}) - \sin 2\phi^{\prime}\
{\cal U}(\bi{\theta}^{\prime})$ and ${\cal U}_r=\cos 2\phi^{\prime}\
{\cal U}(\bi{\theta}^{\prime}) + \sin 2\phi^{\prime}\
{\cal Q}(\bi{\theta}^{\prime})$. The window is $\omega(\theta)=-1/\pi\theta^2\;
(\theta\neq 0)$, $\omega(\theta)=0 \;
(\theta= 0)$. 
From  equations (\ref{avgprime})
and  (\ref{expansion}) it follows that 
\begin{eqnarray}
\langle {\cal S} \rangle_{CMB} &=& 2\kappa \nonumber \\
\langle {\cal E} \rangle_{CMB} &=& 2\kappa \nonumber \\
\langle {\cal B}\rangle_{CMB} &=&0. 
\label{recon}
\end{eqnarray}


Equation (\ref{recon}) shows that the average 
of ${\cal S}$
and ${\cal E}$ can be used to reconstruct
the projected mass density. However, in any particular direction on the sky the CMB
derivatives can take any value so there is an intrinsic noise in the
reconstruction coming from the random nature of the CMB. 

To describe the CMB noise 
we need to calculate correlation functions between $\tl {\cal S}$, $\tl
{\cal Q}$ and $\tl {\cal U}$. These can be expressed  
in terms of the correlation functions of the derivatives
of the unlensed CMB field. For simplicity we consider two directions 
separated by an angle $\theta$ in the $x$ direction,
\begin{eqnarray}
C_{xx}(\theta)&\equiv&\langle \tl T_x(0) \tl T_x(\theta) \rangle_{CMB} \nonumber \\
&=&(2\pi)^{-2}\int d^2{\bi l}\ 
e^{i l\cdot \theta \cos\phi_l}\ l^2C^{\tl T \tl T}_l \cos^2\phi_l \nonumber \\
&=&\int {l dl \over 4 \pi}\ 
l^2C^{\tl T \tl T}_l \left[ J_0(l\theta)-J_2(l\theta)\right]\nonumber \\
&\equiv&{1 \over 2}[C_0(\theta)-C_2(\theta)] \nonumber \\
C_{yy}(\theta)&\equiv&\langle \tl T_y(0) \tl T_y(\theta) \rangle_{CMB} \nonumber \\
&=&(2\pi)^{-2}\int d^2{\bi l}\ 
e^{i l\cdot \theta \cos\phi_l}\ l^2C^{\tl T \tl T}_l \sin^2\phi_l \nonumber \\
&=&\int {l dl \over 4 \pi}\ 
l^2C^{\tl T \tl T}_l \left[ J_0(l\theta)+J_2(l\theta)\right] \nonumber \\ 
&\equiv&{1 \over 2}[C_0(\theta)+C_2(\theta)] \nonumber \\
C_{xy}(\theta)&\equiv&\langle \tl T_x(0) \tl T_y(\theta) \rangle_{CMB} \nonumber \\
&=&(2\pi)^{-2}\int d^2{\bi l}\ 
e^{i l\cdot \theta \cos\phi_l}\ l^2C^{\tl T \tl T}_l 
\cos\phi_l \sin\phi_l \nonumber \\
&=& 0,
\label{correlations}
\end{eqnarray} 
where $C_0(\theta)$ and 
$C_2(\theta)$ are defined as 
the integrals over $l^3C_ldl/2\pi$ weighted with $J_0(l \theta)$ and 
$J_2(l\theta)$, respectively.
The real space correlations of $\tl {\cal S}$,  
$\tl {\cal Q}$,  $\tl {\cal U}$ are 
\begin{eqnarray}
N^{\cal S  S}(\theta)&\equiv & 
\langle \tl {\cal S}(0) \tl {\cal S}(\theta)\rangle 
= 2(C_{xx}^2+C_{yy}^2)/ \sigma_{\cal S}^2 = (C_0^2+C_2^2)/\sigma_{\cal S}^2\nonumber \\
N^{\cal  Q  Q}(\theta)&\equiv &
\langle \tl {\cal Q}(0) \tl {\cal Q}(\theta)\rangle 
= 2(C_{xx}^2+C_{yy}^2)/ \sigma_{\cal S}^2 =(C_0^2+C_2^2)/\sigma_{\cal S}^2\nonumber \\
N^{\cal U  U}(\theta)&\equiv &
\langle \tl {\cal U}(0) \tl {\cal U}(\theta)\rangle 
= 4 C_{xx} C_{yy}/ \sigma_{\cal S}^2 =(C_0^2-C_2^2)/\sigma_{\cal S}^2\nonumber \\
N^{\cal  S  Q}(\theta)&\equiv &
\langle \tl {\cal S}(0) \tl {\cal Q}(\theta)\rangle 
=2(C_{xx}^2-C_{yy}^2)/ \sigma_{\cal S}^2 = 2C_0C_2/\sigma_{\cal S}^2
\nonumber \\
N^{\cal S  U}(\theta)&\equiv &
\langle \tl {\cal S}(0) \tl {\cal U}(\theta)\rangle = 0 \nonumber \\
N^{\cal  Q  U}(\theta)&\equiv &
\langle \tl {\cal Q}(0) \tl {\cal U}(\theta)\rangle = 0. 
\label{corrsqu}
\end{eqnarray}
With the normalization we have chosen the correlations for ${\cal S}$,
${\cal Q}$ and ${\cal U}$ 
at zero lag are equal to one.
The correlations when the separation is not along the $x$ axis can be
obtained by rotations of those in equation (\ref{corrsqu}), 
in a similar way as done for
the correlations of the Stokes parameters that describe the CMB
polarization \cite{urospol,kks}. 

To reconstruct the power spectrum of $\kappa$ it will
be necessary to have expressions for the noise power spectra $N_l^{\cal S S}$, 
$N_l^{\cal E E}$
and $N_l^{\cal S E}$, defined as 
\begin{equation}
\langle {\cal W}({\bi l}){\cal W}'({\bi l}) \rangle=
4C_l^{\kappa \kappa}+N_l^{\cal WW'}, 
\label{estim}
\end{equation}
where $\cal W$ stands for ${\cal S}$ or $\cal E$.
For $\cal B$ only the noise term $N_l^{\cal BB}$ contributes. 
These power spectra can be obtained
from the real space correlation functions using \cite{lensps}
\begin{eqnarray}
N^{\cal S S}_l&=&2\pi \int  \theta d\theta \ N^{\cal S S}(\theta)
\ J_0(l\theta) \nonumber \\
N^{\cal E E}_l&=&\pi \int  \theta d\theta 
\{ [N^{\cal Q Q}(\theta)+N^{\cal U U}(\theta)]\ J_0(l\theta)
+ [N^{\cal Q Q}(\theta)-N^{\cal U U}(\theta)]\ J_4(l\theta) \} \nonumber \\
N^{\cal B B}_l&=&\pi \int  \theta d\theta \
\{ [N^{\cal Q Q}(\theta)+N^{\cal U U}(\theta)]\ J_0(l\theta)
- [N^{\cal Q Q}(\theta)-N^{\cal U U}(\theta)]\ J_4(l\theta) \} \nonumber \\
N^{\cal S E }_l&=&2\pi \int  \theta d\theta \
N^{\cal S E }(\theta)\ J_2(l\theta).
\label{defcl}
\end{eqnarray}

From equations (\ref{corrsqu}) and (\ref{defcl}) it follows that 
the CMB noise power spectra are,
\begin{eqnarray}
N^{\cal S S}_l
&=&{2\pi \over \sigma_{\cal S}^2}\int  \theta d\theta 
(C_0^2+C_2^2)J_0(l \theta)\nonumber \\
N^{\cal E E}_l
&=&{2\pi \over \sigma_{\cal S}^2} \int  \theta d\theta [C_0^2J_0(l \theta)+C_2^2
J_4(l \theta)]\nonumber \\
N^{\cal B B}_l
&=&{2\pi \over \sigma_{\cal S}^2} \int  \theta d\theta [C_0^2J_0(l \theta)-C_2^2
J_4(l \theta)]\nonumber \\
N^{\cal S E }_l
&=&{4\pi \over \sigma_{\cal S}^2} \int  \theta d\theta C_0C_2J_2(l \theta).
\label{ps4}
\end{eqnarray}
 
\begin{figure*}
\begin{center}
\leavevmode
\epsfxsize=4.0in \epsfbox{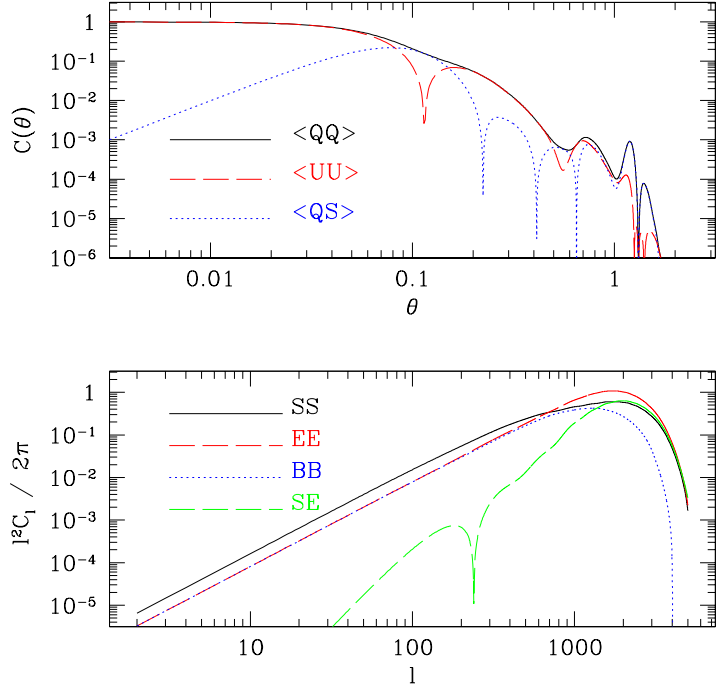}
\end{center}
\caption{The upper panel shows the 
correlation function of $\tl S$, $\tl Q$, $\tl U$ for
SCDM. The lower panel shows the power spectra of $\cal SS$, $\cal EE$, $\cal BB$ and 
$\cal SE$.}
\label{corrder}
\end{figure*}

The top panel of figure \ref{corrder} shows the correlation functions from equation 
(\ref{corrsqu}), while the bottom panel of the same figure shows
the different noise power spectra using
a standard CDM temperature power spectrum. 
The correlations in the CMB derivatives drop significantly for angles larger than 
$\xi=0.15^{\circ}$. 
The power spectra shown in figure \ref{corrder} 
demonstrate that the large scale behavior of the correlations 
is like white noise.
In the limit of low $l$ we can
obtain the power spectrum from equation (\ref{ps4}) by integrating over angle first 
and using the orthonormality relation of the Bessel functions,
\begin{eqnarray}
\lim_{l\rightarrow 0} N_l^{{\cal S}{\cal S}}&=&2\pi\int \theta d\theta
(C_0^2+C_2^2)/\sigma_{\cal S}^2 \nonumber \\ &=&
(2\pi\sigma_{\cal S}^2)^{-1}\int \theta d\theta \int l^3dl \int l'^3dl'C_l^{TT}C_{l'}^{TT}[J_0(l\theta)
J_0(l'\theta)+J_2(l\theta)J_2(l'\theta)] \nonumber \\ 
&=&{ 1\over \pi}\ \ 
{\int l^5dl (C_l^{TT})^2 \over (\int l^2 dlC^{TT}_l)^2} \nonumber \\
\lim_{l\rightarrow 0} N_l^{{\cal E}{\cal E}}&=&2\pi\int \theta d\theta
C_0^2/\sigma_{\cal S}^2 =
\lim_{l\rightarrow 0} N_l^{{\cal B}{\cal B}} 
={1 \over 2} \lim_{l\rightarrow 0} N_l^{{\cal S}{\cal S}}
\nonumber \\
\lim_{l\rightarrow 0} N_l^{{\cal S}{\cal E}} &=& 0. 
\label{limitscl}
\end{eqnarray}
We can use these to 
define the correlation length more precisely as 
$\xi^2= N_l^{\cal SS}$ which gives $\xi\sim 0.15^{\circ}$ used above. 
It is interesting to note that although ${\cal S}$,  ${\cal Q}$,
${\cal U}$ all have the same variance the power spectrum 
of $\cal S$ is twice that of $\cal E$ or $\cal B$. 
$\cal E$ and $\cal B$ have a different
correlation length from $\cal S$, which reflects the spin nature of the shear variables.    
This leads to $\cal E$ giving a factor of 2 higher signal to noise than $\cal S$ in the 
reconstruction.

In order to get an accurate measurement of the
projected mass density or the shear we need to average over several
``independent'' patches of size $\xi^2$, 
each of which has a variance of order unity. This 
sets the basic requirement that has to be satisfied if the reconstruction is 
to give a positive signature.
The signal to noise in each patch of size $\xi^2$
is $ S/N \sim 2\kappa$.
This means that we need to average over $N_{patch}^{-1/2} \sim
2 |\kappa|$ to get a signal to noise of order one. The issue 
is whether on scales larger than $\xi$ the weak lensing signal
is sufficiently strong to be detectable. Even though CMB  
is sensitive to matter density fluctuations up to $z \sim 1100$ and 
the RMS convergence is significantly higher than for 
galaxies at say $z \sim 1$, it is still well below the CMB noise if 
$\xi=0.15^{\circ}$.
Only with sufficient 
small scale power can individual structures be reconstructed with high 
enough signal to noise. This is discussed further in the following sections. 

Equations (\ref{ps4}) are needed to asses the noise for the
reconstruction
of $\kappa$ and to subtract the noise contribution to its
reconstructed
power spectrum in equation (\ref{estim}). 
To compute the variance  of the power spectrum reconstruction
we also need to know the RMS of the CMB noise power spectrum. 
The CMB noise is a fourth order statistics of CMB field. Fortunately in
the large scale limit it can be considered Gaussian.
This is a consequence of the
central limit theorem: the long wavelength modes of ${\cal S}$, ${\cal E}$ and
${\cal B}$ are obtained by adding a lot of independent
patches making them effectively Gaussian. 

If the CMB noise can be considered
Gaussian then the covariance matrix ${\rm Cov}(XX')$ for the noise power spectra can be
expressed in terms of these \cite{spinlong},
\begin{eqnarray}
{\rm Cov}[(N^{\cal S S}_l)^2] &=& {2\over 2l+1} (N^{\cal S S}_l)^2 \nonumber \\
{\rm Cov}[(N^{\cal E E}_l)^2] &=& {2\over 2l+1} (N^{\cal E E}_l)^2 \nonumber \\
{\rm Cov}[(N^{\cal S E}_l)^2] &=& {1\over 2l+1} [(N^{\cal S E}_l)^2+N^{\cal S
S}_l N^{\cal E E}_l] \nonumber \\
{\rm Cov}(N^{\cal S S}_l N^{\cal E E}_l) &=& {2\over 2l+1} 
(N^{\cal S E}_l)^2 \nonumber \\
{\rm Cov}(N^{\cal S S}_l N^{\cal S E}_l) &=& {2\over 2l+1} N^{\cal S E}_l 
N^{\cal S S}_l \nonumber \\
{\rm Cov}(N^{\cal E E}_l N^{\cal S E}_l) &=& {2\over 2l+1} N^{\cal S E}_l 
N^{\cal E E}_l .
\label{covariances}
\end{eqnarray}

\begin{figure*}
\begin{center}
\leavevmode
\epsfxsize=4in \epsfbox{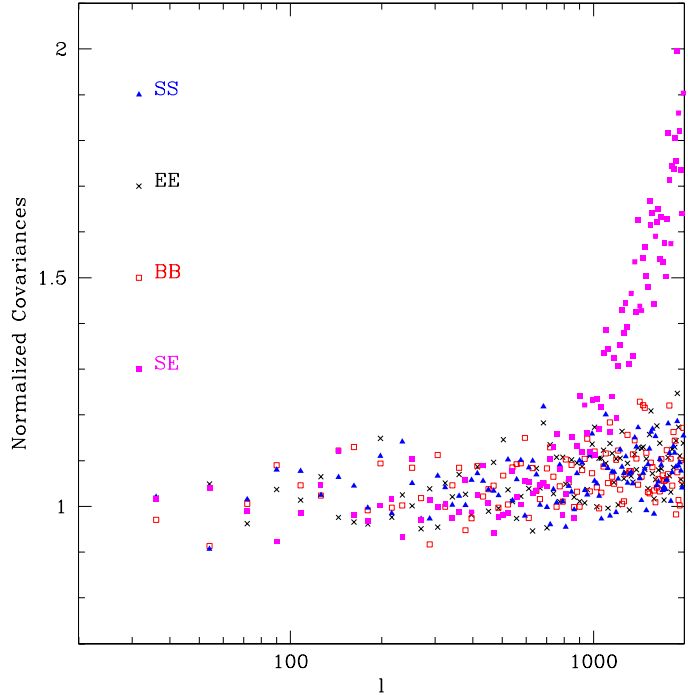}
\end{center}
\caption{Simulation results for the covariances 
of $N_l^{\cal S S}$, $N_l^{\cal E E}$ $N_l^{\cal
B B}$ and $N_l^{\cal S E}$ normalized to their values for Gaussian
noise given in equation (\ref{covariances}). The Gaussian approximation is
an excellent one for $l<1000$.}
\label{scatter}
\end{figure*}

In the large scale limit
$ N^{\cal S E}_l \ll N^{\cal E E}_l$ 
and $N^{\cal S S}_l=2N^{\cal E
E}_l=2N^{\cal B B}_l$. The covariance matrix becomes diagonal with 
${\rm Cov}[(N^{\cal E E}_l)^2]={\rm Cov}[(N^{\cal B B}_l)^2]={\rm Cov}[(N^{\cal S
E}_l)^2]={\rm Cov}[(N^{\cal S S}_l)^2]/4$. 
In figure \ref{scatter} 
the covariances obtained in the simulations are compared to those 
derived under the Gaussian approximation. 
The agreement is excellent for $l<1000$ 
and only seriously breaks down beyond that for $N^{\cal SE}$.
We have also verified with Monte Carlo simulations that the 
off-diagonal terms are negligible compared to the diagonal ones.

In equation (\ref{estim}) we have three 
estimators for the convergence power spectrum,  
${\cal S S}$, ${\cal EE }$ and the cross correlation ${\cal SE }$. 
Using the covariance matrix in (\ref{covariances}) we can
construct a  minimum variance
combination of the three estimators. In the large scale limit
the covariance matrix is diagonal and we can 
weight each of the estimators by the inverse of noise variance,
\begin{eqnarray}
4\hat C^{\kappa \kappa}_l&=&{1 \over 9}(C_l^{\cal S \cal S}-N_l^{\cal
S \cal S})+ {4 \over 9}(C_l^{\cal E \cal E}-N_l^{\cal E \cal E})+ {4
\over 9}(C_l^{\cal S \cal E}-N_l^{\cal S \cal E}) 
\label{waverage}
\end{eqnarray}

Note that we have assumed in equation (\ref{covariances}) 
that only the CMB noise 
contributes to the variance of $C_l^{\kappa \kappa}$ estimates.
There is also the cosmic variance contribution 
${2 \over 2l+1}(C_l^{\kappa \kappa})^2$ but we will be show in the following sections
that this term
is significantly smaller than the CMB noise term and
can be ignored, unless the CMB has more
small scale power than expected within the currently popular cosmological 
models.

\subsection{Small Scale Limit}

So far we have described the gravitational lensing effect on large scales compared to the
correlation length of the CMB. In this limit we average over (almost)
independent patches of CMB fluctuations, which is similar to the procedure
used in weak lensing, where we average over independent ellipticities of
background galaxies. We now turn to scales much smaller than the
correlation length $\xi$. 
In this limit we can take the derivatives 
$\tl T_x$ and $\tl T_y$ to be constant across the field.  
The physical picture is very different from
the one discussed in previous subsection. 
Weak lensing in this limit
acts as a generator of small scale power. To see this we can imagine the effect of
a small clump of mass on a pure gradient of temperature field. A mass clump will 
magnify and stretch a small patch, which will change the slope of the gradient at 
the position of the clump. The resulting temperature 
field is no longer a pure gradient, but has a small wiggle
superimposed on it.
Small scale power has been 
generated only where the gradient of primary anisotropies is
non-zero because where the surface 
brightness is constant its conservation requires that it remains so even in the 
presence of lensing. 

We now
attempt to use this physical picture to reconstruct the weak 
lensing signal.
Introducing $a=T_x/\sigma_{\cal S}$ and $b=T_y/\sigma_{\cal S}$ we have
\begin{eqnarray}
{\cal S}&=&(2\kappa-1)(a^2+b^2)+2\gamma_1(a^2-b^2)+2\gamma_2 \ 2 a b
\nonumber \\
{\cal Q}&=&(2\kappa-1)(a^2-b^2)+2\gamma_1(a^2+b^2)
\nonumber \\
{\cal U}&=&(2\kappa-1)2 a b+2\gamma_2(a^2+b^2).
\label{squssl}
\end{eqnarray}
We did not subtract unity out of the definition of $S$ as in 
equation (\ref{redef}), because
even though on average 
$\langle (a^2+b^2) \rangle_{CMB}=1$, we are considering a very
small field over which $a$ and $b$ are approximately constant
with $(a^2+b^2)\neq 1$ in general.

We want to determine $a$, $b$, $\kappa$, $\gamma_1$ 
and $\gamma_2$. If we assume that the mean value of 
$\kappa$, $\gamma_1$ and $\gamma_2$ over the field is zero then
\begin{eqnarray}
\langle {\cal S} \rangle &=& a^2+b^2 \nonumber \\
\langle {\cal Q} \rangle &=& a^2-b^2 \nonumber \\
\langle {\cal U} \rangle &=& 2 a b,
\label{consistency}
\end{eqnarray}
where the mean value is taken over a region large enough that the mean 
shear and projected mass density vanish but over which the CMB
gradient remains nearly constant. In practice this can be achieved by 
filtering out the small scale power, so that the remaining power is
largely dominated by primary anisotropies. We can then compute $a$ and
$b$ and use them in equation (\ref{consistency}).

Once $a$ and $b$ are determined it would appear that we can determine
the convergence and the two shear components
by solving 
\begin{equation}
\left(\begin{array}{c}
{\cal S} \\ {\cal Q} \\ {\cal U} 
\end{array}\right)=
\left(
\begin{array}{c c c}
\langle {\cal S} \rangle & \langle {\cal Q} \rangle &  \langle {\cal U} \rangle \\ 
\langle {\cal Q} \rangle & \langle {\cal S} \rangle &  0 \\ 
\langle 0 \rangle & \langle {\cal U} \rangle &  \langle {\cal S} \rangle \\ 
\end{array}\right) \
\left(
\begin{array}{c}
1-2\kappa \\ -2 \gamma_1 \\ -2\gamma_2 
\end{array}\right).
\end{equation}
However, the determinant of the matrix above vanishes so we cannot 
determine convergence and shear independently. Instead we have to use
the relations between shear and convergence to reconstruct the 
signal. In Fourier space,
for $l$ large enough that
we can consider $a$ and $b$ in equation (\ref{squssl}) as constants.  We get,
\begin{eqnarray}
{\cal S}(\bi{l})&=& 2 \kappa({\bi l}) \left[
(a^2+b^2)+(a^2-b^2)\cos(2\phi_{\bi{l}})+2a b \sin(2\phi_{\bi{l}})
\right] \nonumber \\
{\cal Q}(\bi{l})&=&2 \kappa(\bi{l})\left[
(a^2-b^2)+(a^2+b^2)\cos(2\phi_{\bi{l}})\right]
\nonumber \\
{\cal U}(\bi{l})&=&2 \kappa(\bi{l})\left[
2 a b+(a^2+b^2)\sin(2\phi_{\bi{l}}) \right].
\label{convsl}
\end{eqnarray} 
We can compute ${\cal E}$ and ${\cal B}$,
\begin{eqnarray}
{\cal E}(\bi{l})&\equiv&{\cal Q}(\bi{l})\cos(2\phi_{\bi{l}})+{\cal
U}(\bi{l}) \sin(2\phi_{\bi{l}})
\nonumber \\
&=&2 \kappa(\bi{l})\left[
(a^2+b^2)+(a^2-b^2)\cos(2\phi_{\bi{l}})+2a b \sin(2\phi_{\bi{l}})
\right] \nonumber \\
&=&{\cal S}(\bi{l}) \nonumber \\
{\cal B}(\bi{l})&\equiv&{\cal U}(\bi{l})\cos(2\phi_{\bi{l}})-
{\cal Q}(\bi{l})\sin(2\phi_{\bi{l}})
\nonumber \\
&=&2 \kappa(\bi{l})\left[
2a b \cos(2\phi_{\bi{l}})+ (a^2-b^2) \sin(2\phi_{\bi{l}})
\right]. 
\end{eqnarray}
We see that 
the estimators of the
convergence are not independent, ${\cal E}(\bi{l})={\cal S}(\bi{l})$, 
as argued above.

To obtain an estimator of convergence we can use any of the expressions
in equation (\ref{convsl}). This requires dividing with the combination of 
derivatives of $T$ and will be very noisy if both $a$ 
and $b$ are close to 0. This is a consequence of the fact that 
lensing cannot generate power where there is no gradient.  
Such a reconstruction will therefore have variable noise. One solution
to this problem is to filter the map with a variable filtering length.
Note that we have assumed $a$ and $b$ are constant 
across the map, while in reality they will change as well. This means
that in practice we have to divide the map into chunks over which the
long wavelength modes are approximately constant, or use more sophisticated
methods such as the wavelet analysis. 

The procedure outlined above is quite involved and we did not attempt
to implement it in this paper. It simplifies considerably
if we are only interested in the power spectrum. In this case we may still use  
${\cal S}(\bi{l})$ and ${\cal B}(\bi{l})$ to get,
\begin{eqnarray}
C^{\cal S S }_{\bi l} &=&4 C^{\kappa \kappa}_l\left[(a^2+b^2)^2
+(a^2-b^2)^2 \cos^2(2\phi_{\bi{l}}) 
+(2ab)^2 \sin^2(2\phi_{\bi{l}})\right. \nonumber \\
&+& \left. 2 (a^2+b^2) (a^2-b^2) \cos(2\phi_{\bi{l}})+
4 (a^2+b^2)ab \sin(2\phi_{\bi{l}})+4 (a^2-b^2) ab \cos(2\phi_{\bi{l}})
\sin(2\phi_{\bi{l}})  \right] \nonumber \\
C^{\cal B B }_{\bi l}&=&4C^{\kappa \kappa}_l\left[
+(2ab)^2 \cos^2(2\phi_{\bi{l}}) 
+(a^2-b^2)^2  \sin^2(2\phi_{\bi{l}})\right. \nonumber \\
&-& \left. 4 (a^2-b^2) ab \cos(2\phi_{\bi{l}})
\sin(2\phi_{\bi{l}})  \right].
\end{eqnarray}
If we average over $\phi_{\bi{l}}$ and use $\langle (a^2+b^2)^2
\rangle_{CMB}=\langle \tl T_x^4 + \tl T_y^4 + 2 \tl T_x^2 \tl T_y^2
\rangle_{CMB}/\sigma_{\cal S}^2 = 2$ and $\langle (a^2-b^2)^2
\rangle_{CMB}=\langle (2 a b)^2 \rangle_{CMB}=1$ we find,
\begin{eqnarray}
C^{\cal S S }_l &=& 12  C^{\kappa \kappa}_l \nonumber \\ 
C^{\cal BB }_l &=& 4 C^{\kappa \kappa}_l. 
\label{factor3}
\end{eqnarray} 
Note that
there is no need to use local estimates of $a$ and $b$ to get an 
estimator of the power spectrum, as long we have a CMB map with enough
uncorrelated patches so that the quadratic combinations of $a$ and $b$
can be replaced by their averages. In this limit
the power spectra of ${\cal S}$ and ${\cal B}$ again
directly give estimates of 
convergence power spectrum, without any convolution and also without
noise contribution from 
intrinsic CMB anisotropies (although there will be contribution from 
instrumental noise). This is because in this limit all the power is generated
by gravitational lensing.
It is important 
to note that in this limit $\cal B$ does not vanish, but actually gives 
an estimate of the $\kappa$ power spectrum, while power in ${\cal S}$
is a factor of $3$ bigger than power in ${\cal B}$.
We will show below that these predictions are well recovered in the
Monte Carlo simulations.

\section{Reconstructing the mass profile of a cluster}

To illustrate our method we first apply it to reconstruct the mass profile
of a massive cluster of galaxies. In order for the large scale limit method
to work we need the CMB to vary on scales smaller than the cluster, so
that the effect of lensing can be measured after averaging over independent
patches.
Since we do not expect the primary CMB anisotropies to have
sufficient power on arcminute scales we use 
the Ostriker-Vishniac (OV) effect instead in our example
using the power spectrum from \cite{jaffekam} as an estimate. 
This is not necessary the only possible
source of anisotropies on small scales and there may be other sources 
as well, including the primeval galaxies and QSO emitting in 
IR and radio. As long as there are many sources
in the beam and their redshift is higher than the cluster redshift 
the analysis is identical to the one presented here. 
The only difference is that these sources are not at $z \sim 1100$ but at 
lower redshift
so the window function $g$ in equation (\ref{kappa}) changes relative to the 
one for $z \sim 1100$ sources.
If the redshift distribution of the sources is 
not known then we may in principle use 
lensing on a cluster with known mass profile
to determine it.
For individual sources that are resolved it may be better to use their shapes
as an estimate of the local shear, which is the usual procedure in the case
of galaxy lensing.

For the method to be useful in practice 
the small scale anisotropies have to be above the detector noise. This could 
be achievable with future interferometric or bolometer experiments with 
long integration times on a small area of the sky, reaching $\mu K$ sensitivities
on subarcminute pixels. Another complication in
this case relative to our analysis in \S 2 is that the distribution of
anisotropies is likely not to be Gaussian. This complicates the issue
of noise analysis and statistical significance of the results, but does 
not affect the average of the reconstruction. 
We will ignore these issues in the following, since we are mainly 
interested in having a simple
example with which we can test our method. 

\begin{figure*}
\begin{center}
\leavevmode
\epsfxsize=6in \epsfbox{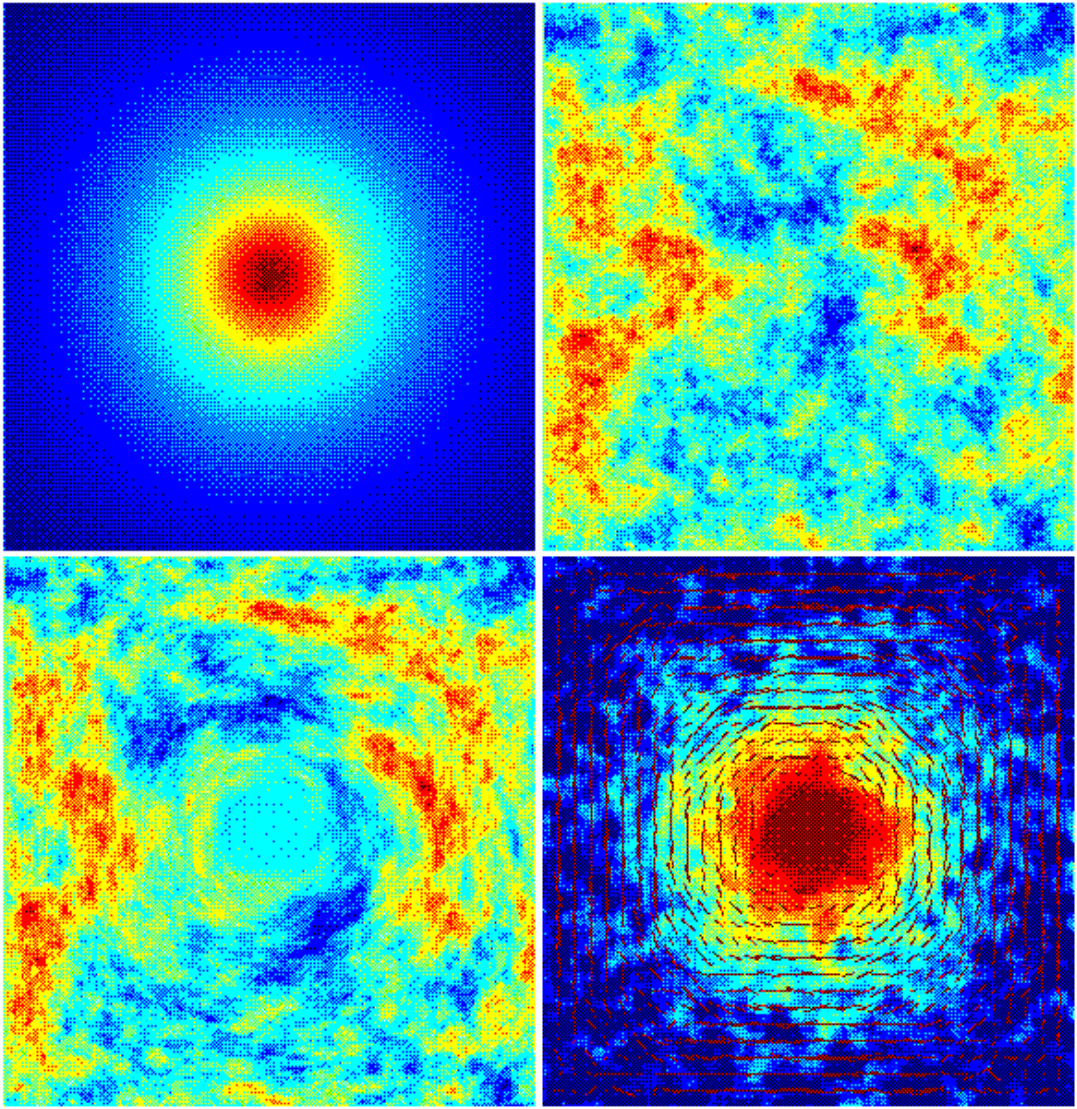}
\end{center}
\caption{
Top left panel: input cluster on a 6'$\times$6' field. Top right panel: unlensed CMB map. 
We assumed
that the Ostriker-Vishniac effect could be detected with sufficient signal
to noise to be used in the reconstruction of $\kappa$. Bottom left: 
CMB field after being lensed by the cluster. Bottom right: the
background shows the $\cal S$ field while the rods represent the shear
variables ${\cal Q}$ and ${\cal U}$, both of which can be used to reconstruct the 
density profile.}
\label{shear}
\end{figure*}

Figure \ref{shear}a shows the simulated input cluster, with the profile chosen 
as a projection of a truncated isothermal sphere. 
In figure \ref{shear}b we show 
an unlensed  simulated CMB field, assuming OV type of power spectrum. In 
figure \ref{shear}c we show the
lensed CMB map. We see that the cluster has magnified the CMB 
in the center, so in the absence of small scale power the central region is 
smoother. 
In figure \ref{shear}d we show the shear as measured by the
CMB derivatives ($\cal Q$ and $\cal U$) and in the background the
convergence field $\cal S$. 

Both $\cal S$ and $\cal E$ are estimators of the convergence, but
figure \ref{shear}c illustrates a possible advantage of the reconstruction based on
shear over that based on ${\cal S}$ in the case of clusters. 
The shear based reconstruction is nonlocal as shown from equation (\ref{ebreal}). 
The shear from the whole map is used to reconstruct the density
at the center. Thus we are averaging over many different independent
CMB regions.
On the other hand the reconstruction based on ${\cal S}$ is local. It will
not work very well in the center because the magnification induced by
the lensing will expand a small region so unless there are 
additional small scale perturbations which are magnified 
there will be fewer independent regions to average over. This effect
can be seen in
figure \ref{shear}c. 
This is of course
an issue only for clusters that are close to being critical and the magnification is no
longer small compared to unity. For other, more linear structures the 
two reconstructions give similar results, although $\cal S$ is still
noisier than $\cal E$ by a factor of 2 as derived in equation (\ref{limitscl}). 

\begin{figure*}
\begin{center}
\leavevmode
\epsfxsize=4in \epsfbox{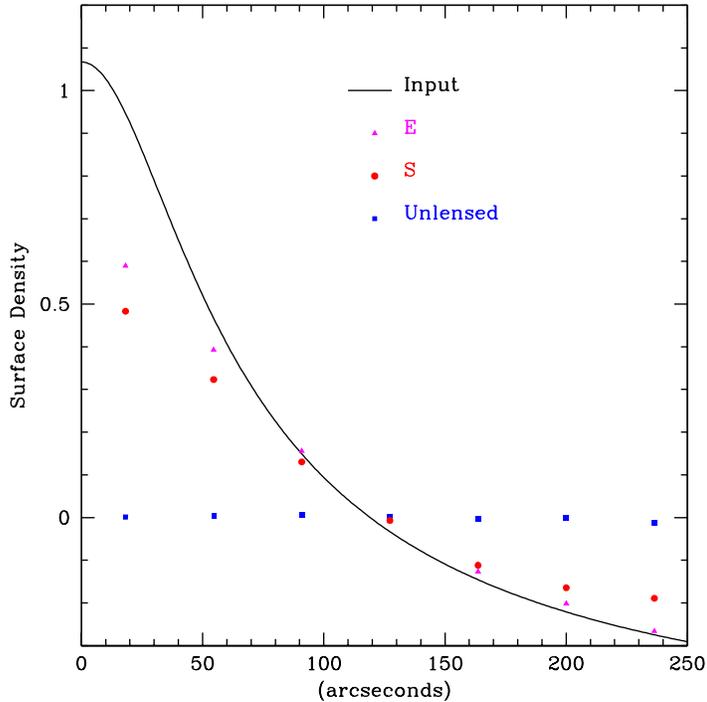}
\end{center}
\caption{Reconstructed radial profile using the data in Figure 2. The
points marked with ${\cal E}$ (${\cal S}$) correspond to the reconstruction of
$\kappa$ based on ${\cal E}$ (${\cal S}$). The points marked unlensed correspond to
the result of applying the ${\cal E}$ reconstruction to the unlensed
field. The input $\kappa$ was taken to have zero mean.}
\label{radial}
\end{figure*}

To be more quantitative we show in 
figure \ref{radial} the reconstructed projected radial
mass profile from the shear and the convergence. 
We also show the result of the reconstruction procedure
on the unlensed CMB field to illustrate the noise level, which 
for this case with a lot of small scale structure is negligible.
The reconstructed profile tracks well the input profile outside the center, 
but falls below it in the center because of insufficient number of independent
patches there. 
In our reconstruction algorithm we are forcing the mean $\kappa$ to be
zero and
because the reconstruction of the
center of the cluster is systematically lower due to the averaging,
the reconstruction is also systematically
lower on the outside of the cluster. 
As expected this is more pronounced for ${\cal S}$ than for ${\cal E}$
reconstruction because the former is noisier and 
relies  more on poor quality information from the center of the cluster.
These effects can be 
calibrated from the simulations and overall the method can reconstruct the 
true projected mass density outside the core, 
assuming that the necessary conditions discussed above (measurable 
primary anisotropies on scales smaller than the typical cluster scale)
are satisfied. 

\section{Reconstructing the power spectrum of projected density}

\subsection{CMB Noise}

\begin{figure*}
\begin{center}
\leavevmode
\epsfxsize=4in \epsfbox{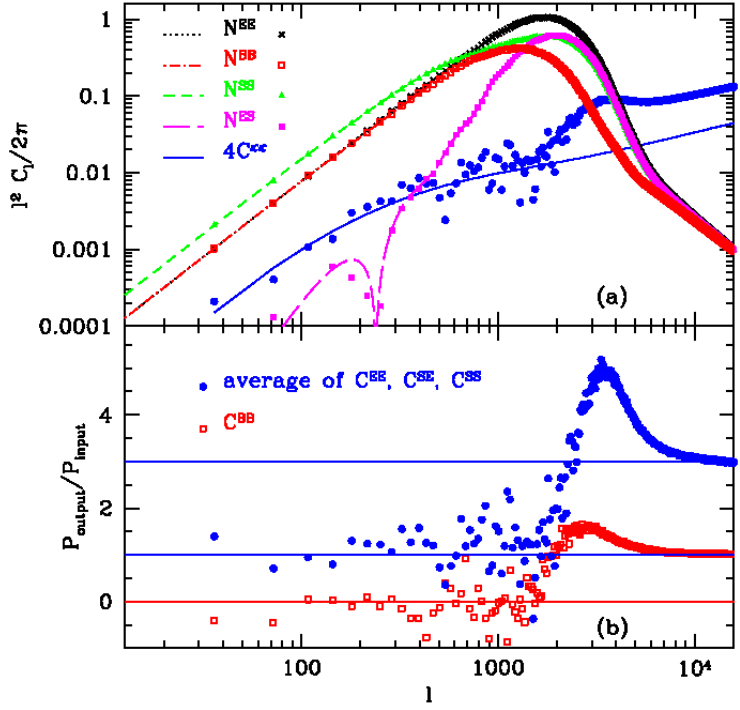}
\end{center}
\caption{(a) $N^{\cal S S}_l$, $N^{\cal E E}_l$, 
$N^{\cal B B}_l$ and $N^{\cal S E}_l$ power spectra
for the unlensed CMB field. The lines correspond to the results of 
equation (\ref{ps4}). Also shown  is the weighted average of the recovered power
spectra from ${\cal S}$, ${\cal E}$ and their cross correlation
together with  the
input $4 C^{\kappa \kappa}_l$. 
(b) Ratio of the 
recovered power spectra to the input $C^{\kappa \kappa}_l$. The signal in
${\cal B}$ is also shown.}   
\label{power1}
\end{figure*}

In this section we investigate the power spectrum reconstruction 
of convergence
using the method developed in \S 2. The power spectrum can be reconstructed  
from ${\cal E}$,  ${\cal S}$ or their cross correlation. 
We will first discuss the power spectrum reconstruction in the absence 
of detector noise and assuming infinite angular
resolution.  In this simplified case the only source of noise is 
the CMB itself. 
We do not add small scale 
secondary anisotropies in this section, so the results should 
be representative for primary fluctuations and any additional 
small scale fluctuations at high redshift would further improve the performance.
In figure \ref{power1}a we show the power spectra for the unlensed CMB
field (the CMB ``noise'') calculated using
equation (\ref{ps4}) and the result of our Monte Carlo simulations. The
agreement between analytic prediction and simulations 
is very good. We recover the white noise behavior 
of $N^{\cal S S}_l$,$N^{\cal E E}_l$ and $N^{\cal B B}_l$ on large
scales and also the relatively 
small amplitude of the cross correlation $N^{\cal S E}_l$. 

On large scales these power spectra are to be compared to $4C_l^{\kappa \kappa}$
(figure \ref{power1}a). For the
purpose of illustrating our  method  in this paper we adopted the convergence power
spectra of the ``concordance'' model \cite{os} $\Omega_m=0.3$,
$\Omega_{\Lambda}=0.7$, $\Gamma=\Omega_m h=0.2$, $n=1$ and
$\sigma_8=1$. 
It can be seen from figure \ref{power1}a that on large scales 
the power of the convergence is small compared to that of the CMB
noise. As a result we can only hope for a statistical detection of
the lensing effect and the method 
will not allow to make a map of $\kappa$ directly unless there is more 
small scale power in the CMB, 
as discussed in \S 3 in the context of cluster reconstruction. 
Even if $\kappa$ reconstruction
has a low signal to noise its 
power spectrum may nevertheless be recovered with high statistical 
significance, 
because we can average over many independent patches on the sky. 
This is seen from the difference between the power spectra in the
lensed and unlensed field. 
The weighted average of the three estimators (equation \ref{waverage}) is
also shown in figure \ref{power1}a together with the input convergence
spectrum. The intrinsic CMB noise is about 10-20 times
higher, yet the difference agrees well with the
input $\kappa$ power spectrum, showing that
there is a significant statistical 
detection of the signal.

The reconstruction of the power spectrum is further studied 
in figure \ref{power1}b, where we show
the CMB noise subtracted power spectra of ${\cal S S}$, 
${\cal E E}$ and ${\cal BB}$ and the cross correlation spectrum ${\cal SE}$
divided with the input matter power spectrum.
On large scales the reconstructed spectra of 
$\cal SS$, $\cal EE$ and $\cal SE$ all give unbiased estimators of $4C_l^{\kappa \kappa}$ 
so the average of the output to input ratio is 1, while 
that of $\cal BB$
if consistent with the pure noise and the ratio averages to 0. 
The latter, although not giving an estimate of the signal, 
can provide an important
consistency check on whether the signal seen from ${\cal E}$ and ${\cal S}$ is 
indeed of cosmological origin or not. It may also help identify the cosmological 
part of the signal if additional contributions are  present, such as in the 
case of non Gaussian CMB fluctuations or contamination from
foregrounds. 

The large scale limit is valid up to 
$l \sim 1000$. 
On very small scales, roughly $l \sim 5000$ and beyond,
the reconstructed power spectra from ${\cal E}$
and ${\cal S}$ give
$12C^{\kappa \kappa}_l$, while that obtained from ${\cal
B}$ gives $4C^{\kappa \kappa}_l$ so that the ratios are 3 and 1, 
again in agreement with analytic prediction 
in equation (\ref{factor3}). 
At very high $l$ CMB noise is negligible, because 
there is no power present in the CMB itself on those scales. 

In the intermediate regime between $l=1000-5000$ neither small scale nor large 
scale limits apply and the reconstruction becomes a complicated convolution 
of CMB anisotropies and lensing signal. 
Information on the power spectrum can 
still be obtained even from this regime, but it is better to  
approach the reconstruction in a parametric form, by parameterizing the 
power spectrum with a few free parameters that can be estimated by fitting the 
simulated spectra to the data, rather than by direct inversion. 

Because CMB noise is larger than the signal on large scales we may worry that a small
error in the noise estimate would lead to a large error in the estimated
power spectrum. This is most worrisome for the 
${\cal S S}$ reconstruction and least 
important 
for the cross correlation
${\cal S E}$ because $N_l^{\cal S S}=2 N_l^{\cal E E} \gg N_l^{\cal S E} $. The latter 
is even smaller than $4C_l^{\kappa \kappa}$ on large scales so we can
obtain an accurate estimate even without any noise subtraction.  
Fortunately the large scale behavior of 
$N_l^{\cal S S}$, $N_l^{\cal E E}$ $N_l^{\cal
B B}$ and $N_l^{\cal S E}$ is quite insensitive to the details of the
power temperature spectra. For example we may worry that 
the CMB power spectra we 
measure will be affected by lensing or that 
we can only know 
the lensed CMB temperature spectra with
some minimum scatter, limited by cosmic variance. On large scales, these effects change  
the mean of $N_l^{\cal E E}$ by approximately 0.5 \% and add a 1\%
scatter to it. It will depend on the noise amplitude whether this is 
a significant source of error. 
For MAP these errors causes a shift in the amplitude
comparable to the input power spectrum, while for Planck it is significantly below it
and can be ignored. For MAP this problem can be solved by using 
$\cal B$ reconstruction. 
Because the effect on large scales is mostly an amplitude shift
we may use the amplitude of $C^{\cal BB}_l$
to make it consistent with 0, which would also properly
determine the amplitude of the 
other spectra.

\subsection{Finite Angular Resolution and Detector Noise}

Next we consider the influence of finite angular resolution and
detector noise. In order to measure the convergence power spectra using the small
scale limit the experiment must have enough angular resolution to
probe these small scales. This effect of finite angular resolution is straightforward
to include and we will not discuss it further, because future CMB satellites such as 
MAP and Planck will not 
be able to probe this limit. Here we discuss the more important
effect of finite angular resolution and detector noise on the large scale limit reconstruction. 
It is easy
to see that finite angular resolution has an effect on the intrinsic
CMB noise amplitude.
Because of finite resolution and detector noise small scale CMB power 
cannot be resolved and this leads to a larger correlation length $\xi$. This means that 
we have fewer independent patches to average over the CMB noise and the overall level of 
noise power spectrum is higher. 
Finite angular resolution also has a more subtle effect by reducing the 
transfered power. 
To understand it quantitatively we first compute the
derivatives of the temperature field in the presence of some filtering function
$F({\bi \theta})$, which involves the experimental beam and any additional filter we
may want to use in the analysis
\begin{eqnarray}
T_a({\bi \theta})&=&\int F(\bi{\theta}-\bi{\theta}')T_a(\bi{\theta}')d^2{\bi \theta}'\nonumber \\
&=& \int F(\bi{\theta}-\bi{\theta}')(\delta_{ab}+\Phi_{ab}){\tl T}_b(\bi{\theta}'+\delta {\bi 
\theta}')d^2{\bi \theta}'
\nonumber \\ 
&=&(2\pi)^{-2}\int d^2{\bi l}  F(l) T_a({\bi
l}) e^{i{\bi l}\cdot \bi{\theta}} \nonumber \\
&+&(2\pi)^{-4}\int d^2{\bi l} \int d^2{\bi q} F(|\bi l+ \bi q|)T_b({\bi
l}) \Phi_{ab} (\bi q)
e^{i{(\bi l+ \bi q)}\cdot \bi{\theta}}.
\label{filter1}
\end{eqnarray} 
To obtain the last expression 
we expanded $F$, $\Phi_{ab}$ and $T_a$ into a Fourier series and integrated
over angle ${\bi \theta}'$.
The large scale effect of the filter
function can be read from the average over CMB of the square of
equation (\ref{filter1}),
\begin{eqnarray}
\langle T_a T_b\rangle_{CMB}&=& {\delta_{ab}\over 2} \int {ldl \over 2\pi}\
F^2(l) l^2C^{TT}_l \nonumber \\
&+&(2\pi)^{-2}\int d^2{\bi q} \Phi_{ab} (\bi q)
e^{i{\bi q}\cdot \bi{\theta}}\left[ (2\pi)^{-2} \int d^2{\bi l} F(l)F(|\bi
l+ \bi q|) C^{TT}_{l} \right] 
\label{filter2}
\end{eqnarray}
In the absence of filtering the integrals involving $C^{TT}_l$ in both
terms of equation (\ref{filter2}) are identical. This  allows us to
reconstruct $\kappa$ from the second term, while the first term gives
$\sigma_{\cal  S}$ as in equation (\ref{averagecmb}). 
To describe the effect of the beam smearing we 
introduce a window function as the ratio of the two
integrals 
\begin{eqnarray}
W(q)={\int d^2{\bi l} F(l)F(|\bi l+ \bi q|)l^2 C^{TT}_{l} \over
 \int d^2{\bi l} \  F^2(l) l^2C^{TT}_l},
\label{window}
\end{eqnarray}
which is in general less than unity and 
in terms of which we have 
\begin{equation}
\langle {\cal S}({\bi l}) \rangle=\langle {\cal E}({\bi l}) \rangle=
2\kappa({\bi l})W(l).
\end{equation} 

The effect of filtering is important
even for the reconstruction of large scale $\kappa$ modes. This can be understood
by looking at equation (\ref{filter1}).
The shear tensor modulates the amplitude of the temperature
derivatives, so the
information about a particular $\bi q$ mode of $\kappa$ is encoded as
sidebands of the different $\bi l$ modes of the temperature. We
are recovering the information back at the corresponding $\bi q$ by
squaring the field,  which appropriately combines back all the
sidebands from every  $\bi l$ to the correct $\bi q$.
Finite angular resolution is important even for
large scale modes because the information about these modes in encoded
mainly in the sidebands of the small scale temperature modes, which are 
strongly affected by the filtering function. 

\begin{figure*}
\begin{center}
\leavevmode
\epsfxsize=4in \epsfbox{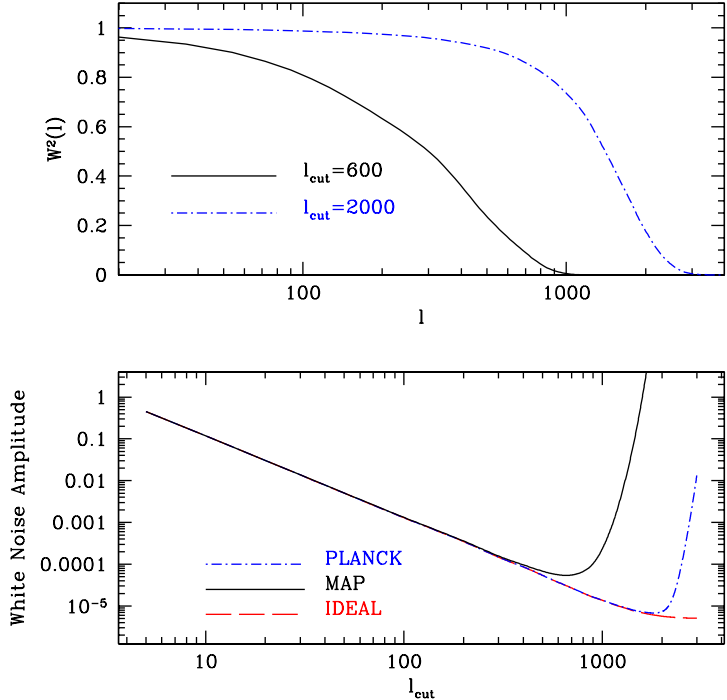}
\end{center}
\caption{The upper panel shows the window function $W^2(l)$ as a function of $l$
for MAP noise and beam properties where the appropriate value is $l_{cut}=600$
and for Planck with $l_{cut}=2000$ (bottom panel). For the latter the window has no effect 
at $l<1000$. Bottom panel shows
amplitude of $N^{\cal E E}_l$ in the limit $l \rightarrow 0$ as a
function of $l_{cut}$ for the specifications of the MAP and Planck satellites, justifying 
$l_{cut}=600, 2000$ for the two experiments. Also shown is the no
noise infinite angular resolution curve (Ideal), which 
levels off at $l \sim 2000$, showing that Planck is close to optimal in the large scale
limit, because it measures most of the power in the CMB.
}
\label{winnoise}
\end{figure*}

To minimize the effect of beam smoothing we may filter the
temperature before squaring it. The smallest effect is achieved if the 
filtering function $F(l)$ is a constant, in which case (in the absence of a cutoff
in $l$) there would be no effect. For this reason we 
choose to filter the temperature with the inverse of the beam $e^{l(l+1)\sigma_b^2}$, 
where $\sigma_b^2$ is the width of the Gaussian beam, 
thereby reversing the smoothing effect of the beam. We can 
only do this up to a maximum $l_{cut}$ (the value of which is determined below)
not to amplify the detector noise on small
scales. Thus our effective filter $F(l)$ 
is equal to one for
$l<l_{cut}$ and zero after that. Figure \ref{winnoise}a shows two
examples of $W(q)$ for different values of $l_{cut}$, which correspond to 
minimum of noise for MAP and Planck experiments. Note that regardless of the 
form of the filter equation (\ref{window}) implies that 
$W(q) \rightarrow 1$ as $q \rightarrow 0$. Other forms of filtering
are possible as well, but we show next that this filter comes close to 
being optimal and is easy to implement in the analysis.

The contribution of detector noise on large scales to the different power
spectra can be obtained 
using equation (\ref{ps4}) with
$C^{TT}_l\rightarrow C_{l}^{TT}+N^{TT}$. 
We take the detector noise to be white noise with constant power spectrum
$N^{TT}$ for a range $0<l< l_{max}$. The power spectrum $N^{TT}$ is related
to the noise level per pixel, $\sigma^2=N^{TT} l_{max}^2/4\pi$ with
$l_{max}^2=N_{pix} 4\pi/\Omega$ and $\sigma$ is expressed in the same units as
$T$ (e.g. $\mu K$). 
If we want the correct normalization for the recovered $C^{\kappa \kappa}_l$
we need to divide by the correct $\sigma_{\cal S}$, which has to be computed
only with the CMB power spectrum, without the detector noise term. With our filtering 
prescription the large scale amplitude
of the $N^{\cal E E}_l$ power spectrum in equation (\ref{limitscl}) becomes
\begin{eqnarray}
N^{\cal E E}_l=2 \pi{\int_0^{lcut} ldl  l^4 F(l)^4(C_{l}^{TT}+e^{l(l+1)\sigma_b^2} 
N^{TT})^2 \over (\int l dl l^2 F^2(l)C^{TT}_{l})^2}.
\label{ncn}
\end{eqnarray}
To derive the optimal filter $F(l)$
we minimize $N^{\cal E E}_l/W^2(l)C^{\kappa \kappa}_l$ with respect to 
$F^2(l)$. In the limit $l \rightarrow 0$ $W(l)=1$ and this becomes equivalent 
to minimizing $N^{\cal E E}_l$ in equation (\ref{ncn}). Taking derivatives 
with respect to $F(l)^2$ and setting to 0 we find
\begin{equation}
F(l)^2={C_l^{TT} \over l^2(C_l^{TT}+e^{l(l+1)\sigma_b^2}
N^{TT})^2}.
\label{opt}
\end{equation}
In the large scale limit this gives $F(l)^2=(l^2C_l^{TT})^{-1}$, which is 
roughly a constant for spectra that are close to scale invariant. 
For large $l$ the noise term dominates over the CMB signal term and 
the filtering function $F(l)$ goes to 0. The transition occurs where 
\begin{equation}
C_l^{TT} \sim e^{l(l+1)\sigma_b^2}
N^{TT}.
\label{lcut}
\end{equation}
To compare this filter to the simple constant filter in 
figure \ref{winnoise}b we show  $N^{\cal E E}_l$ as a function of 
$l_{cut}$ for the constant filter and the 
specifications of noise and angular resolution of
MAP and Planck, as well as in the absence of noise and beam smoothing. 
In the absence of detector noise 
larger $l_{cut}$ is always better because the correlation length is 
reduced, but it saturates beyond $l \sim 2000$ for this model because there
is very little CMB power on small scales. Once detector noise is included   
it eventually dominates at high $l_{cut}$ 
and we are better off not including these modes as they are mostly
noise and do not contribute to the signal.  
In between there is a minimum which determines $l_{cut}$ depending on the 
noise and beam of the experiment. 
The amplitude of noise for this filter
can be compared to
the one in equation (\ref{opt}) which minimizes the noise in the large 
scale limit.
The two give very similar results in the large scale limit. 
Because constant $F(l)$ minimizes the 
effect of the window at higher $l$ 
it is likely to perform even better than the filter in equation (\ref{opt}), 
which was derived under assumption $W(l)=1$. 
For this reason we choose to adopt the simple constant 
filter instead of the one in equation (\ref{opt}).

\subsection{Relation to the Density Field}
\begin{figure*}
\begin{center}
\leavevmode
\epsfxsize=3in \epsfbox{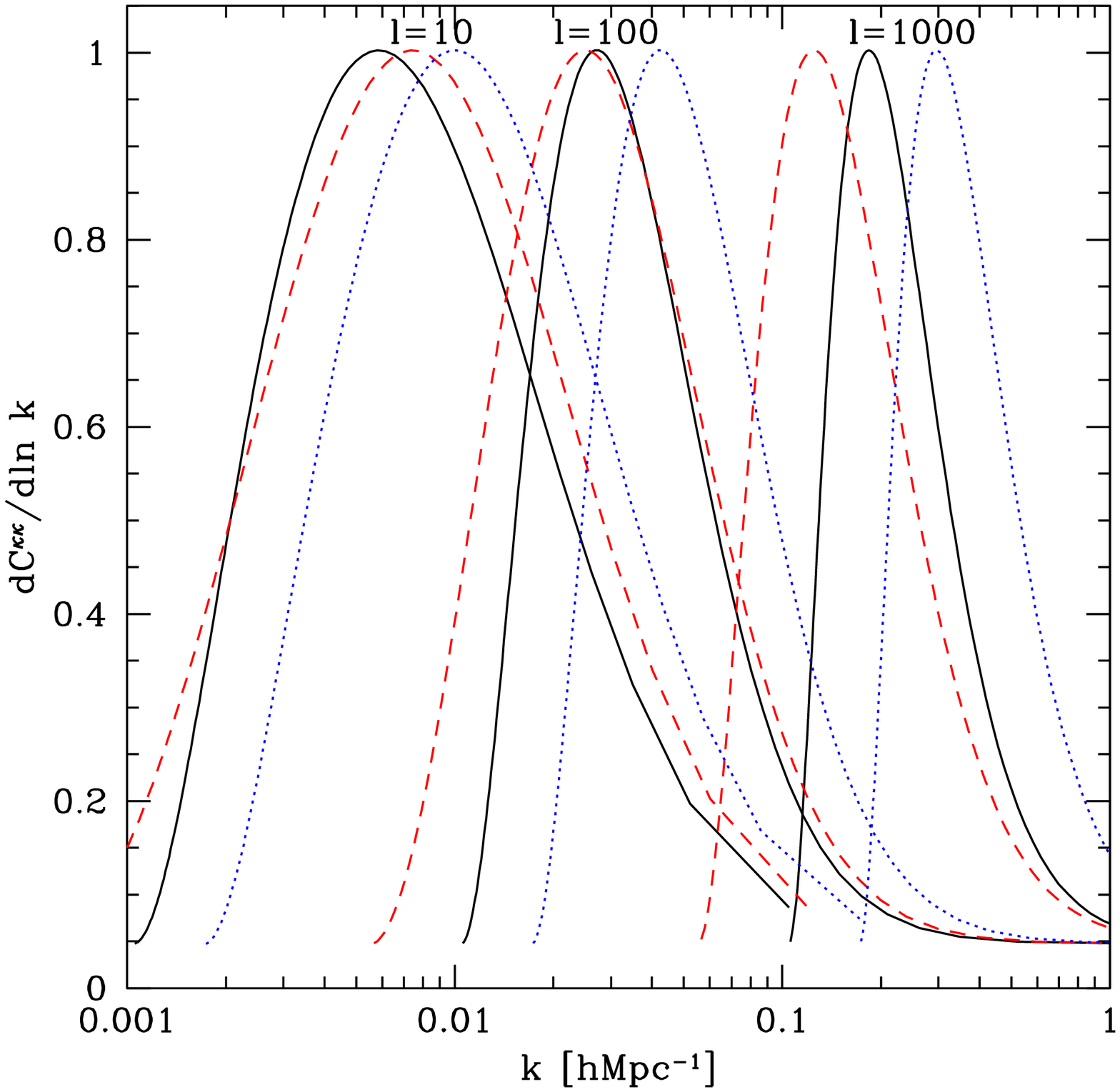}
\epsfxsize=3in \epsfbox{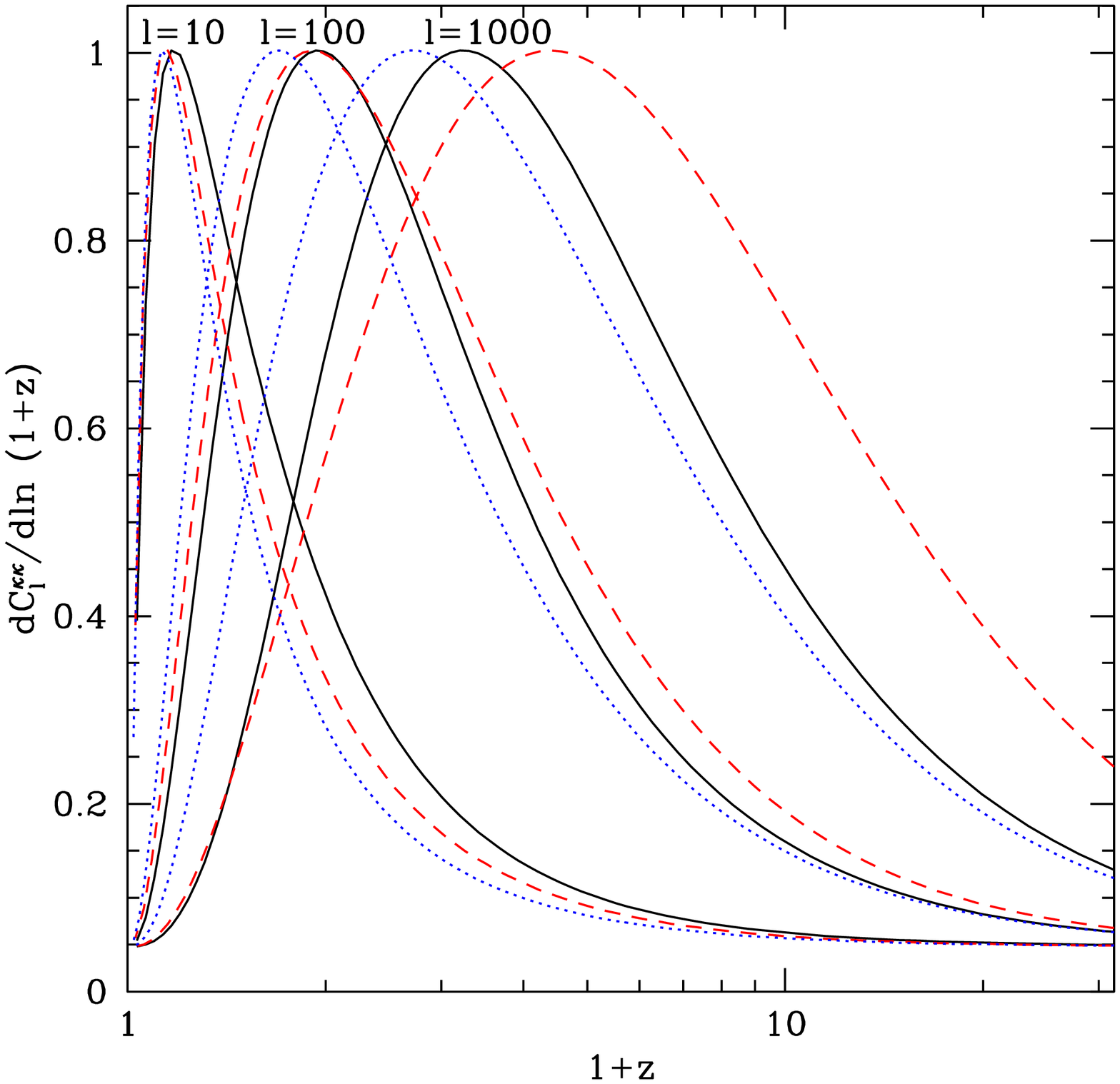}
\end{center}
\caption{Left: logarithmic contribution to $C^{\kappa \kappa}_l$
as a function of $k$ for $l=10$, 100, 1000 (the normalization is
arbitrary).
The models are flat CDM model (dotted), open CDM model with $\Omega_m=0.3$
(dashed) and cosmological constant model with with $\Omega_m=0.3$ (solid). All
the models have $\Gamma=\Omega_m h=0.21$. Right: logarithmic contribution to $C^{\kappa \kappa}_l$
as a function of $1+z$ for the same models as above.
}
\label{fig6}
\end{figure*}

We have shown that the quantity we are  able to extract from the distortions 
of the CMB is the power spectrum of the 
projected density field weighted with 
a window $g/a$ (equation \ref{kappa}). Since the more fundamental quantity is 
the 3-d density field described by the power spectrum we 
would like to know the
relation between the two power spectra. 
This relation is shown in figure \ref{fig6}a, where
a logarithmic
contribution to a given $l$ mode as a function of 3-d wavevector $k$ is plotted for a 
representative sample of models. These windows
are relatively broad functions of $k$ and peak at $\lambda=2\pi/k=1000h^{-1}$Mpc 
for $l=10$ and $\lambda=30 h^{-1}$Mpc for $l=1000$. The exact value  depends on the model.
We are  therefore probing the power spectrum over a range of scales which extends to
scales larger than any other method that directly traces dark matter. 

The next question we want to address is the redshift distribution of the contribution 
to a given $C_l^{\kappa \kappa}$. 
Even though the $g/a$ window peaks at $z=3$ for a 
flat universe this does not mean that the dominant contribution comes from this redshift. 
The relation between $l$ and $k$ depends on the shape of the power
spectrum (equation \ref{cl}).
For any given $l$ there is a range in $k$ that contributes (figure \ref{fig6}a).
The contribution from a high $k$ 
mode comes from structures which are relatively closer to the observer
(and so at lower $z$) 
than the structures that dominate the contribution for a lower $k$
mode (equation \ref{cl}).
On large scales the matter power spectrum has a turnover so low $k$ modes have very little power
which implies that their contribution is smaller than that of 
higher $k$ modes from more nearby structures 
despite the geometrical factor that 
suppresses these higher $k$ modes. Thus
for low $l$ the contribution will be dominated by low $z$ structures. 
The logarithmic
contribution to a given $l$ mode as a function of $1+z$ is shown in figure \ref{fig6}b 
and confirms these expectations. 
At smaller scales the power is more evenly distributed as a function of $k$
and the peak contribution moves to higher redshifts. For $l 
\sim 1000$ it peaks between $z\sim 2-3$ with a long tail extending to higher $z$.
At these scales we are therefore directly probing dark matter 
clustering at high redshifts. 

\begin{figure*}
\begin{center}
\leavevmode
\epsfxsize=4in \epsfbox{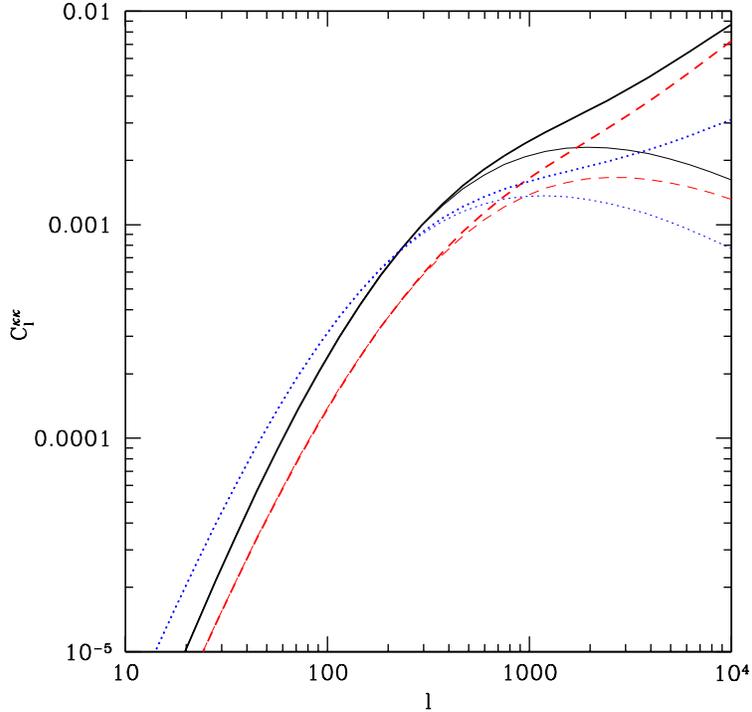}
\end{center}
\caption{Power spectrum $C^{\kappa\kappa}_l$
as a function of $l$ for the same models as in figure \ref{fig6}.
Thin lower curves show the linear power spectra, thick upper curves the nonlinear 
spectra.  }
\label{fig7}
\end{figure*}

The two characteristic features of lensing on the CMB are the large scales 
and early epochs we can probe. In combination these facts  guarantee that the 
power spectrum will be dominated by linear contributions. Figure \ref{fig7}
shows the difference between the $C_l^{\kappa \kappa}$ calculated
using the linear and nonlinear matter power spectrum. The spectra are
shown  as a function of $l$ for the same set of models used in figure 
\ref{fig6}. The nonlinear matter power spectra were computed using the
linear to nonlinear mapping \cite{peacock}. Up to $l \sim 1000$ the 
power spectrum is dominated by linear modes. This has the advantage of
making the results 
simple to interpret in terms of cosmological models, since no nonlinear corrections 
are necessary. Since MAP and Planck reconstruction in the large scale 
limit will not extend beyond
$l \sim 1000$ they will be dominated by linear scales. At higher $l$ nonlinear 
corrections become important. While this has the disadvantage that 
the interpretation becomes more complicated it has 
the advantage in that the power is boosted by almost an order of magnitude 
compared to the linear scales and so the lensing effect becomes more easily observable. 

\section{Cross Correlation with Other Maps}
We have shown that individual structures cannot be reconstructed with sufficient
signal to noise, unless the CMB has more small scale power than expected from primary
anisotropies. One way to obtain a positive detection discussed in previous section 
is by combining information 
from independent patches of the sky into a measurement of the power
spectrum. 
In this section we discuss another 
way to enhance the signal by 
using the cross correlation with another map with higher
signal to noise. A cross correlation
between signal induced by lensing and some other tracer
has been explored before \cite{sss}. The difference to our method is 
that they do not reconstruct $\kappa$ directly, but compute the photon deflection 
angle instead for which the 3-d matter distribution is needed. It can only be 
obtained from a redshift survey (such as SDSS or 2dF), 
under the assumption that light traces mass. Our method can also be used when 
such a 3-d distribution is not available. For example, we may try to 
cross correlate the reconstructed convergence map with
Sunyaev-Zeldovich (SZ) map, which traces the integrated pressure along the line of
sight. This would give positive cross 
correlation because of clusters, which 
should contribute to both convergence and SZ effect. Thus even if clusters 
cannot be individually detected using the method described in \S 3, they may still
be detected statistically through this cross correlation, which would give 
information about pressure to dark matter ratio in clusters.
Other possibilities for cross correlation 
are with X-ray background which is believed to trace large scale structure to $z \sim 5$
and with galaxy catalogs from SDSS and 2dF, both of which would give 
information on how light traces mass on large scales. 
Yet another possibility is the cross 
correlation with the CMB itself, which would give positive detection
whenever there is a significant contribution from the time-dependent potential in the
CMB \cite{isw,ct,spergel}. 
To test the most optimistic possibility we may
estimate the signal to noise of the cross correlation
when we have another perfect map of convergence.  
For any of the probes mentioned above the actual signal to noise will be lower,
since they will only partially correlate with the projected mass density. 
It is nevertheless of interest to see whether such cross correlations would 
be useful at all given such a low signal to noise on individual structures in 
the maps of $\cal S$ and $\cal E$.

\begin{figure*}
\begin{center}
\leavevmode
\epsfxsize=4in \epsfbox{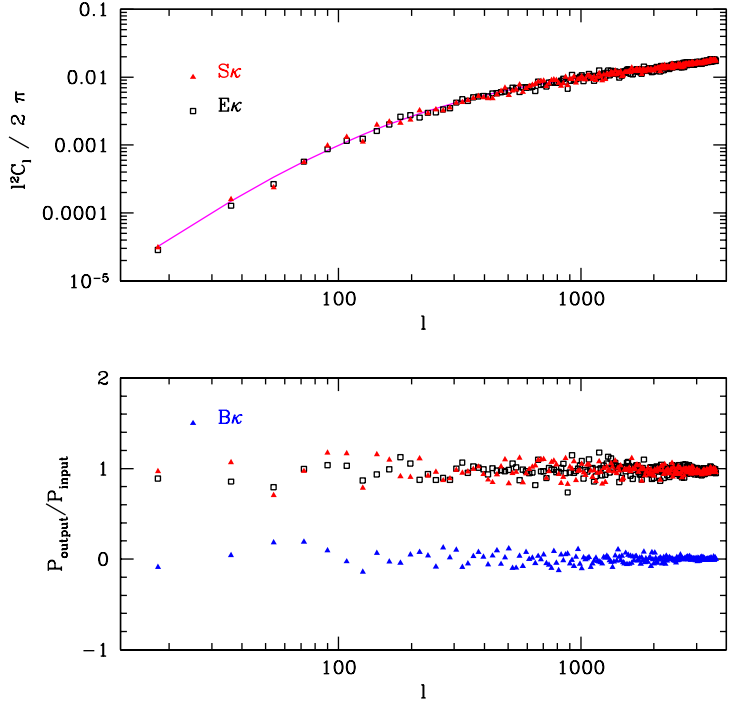}
\end{center}
\caption{Cross correlation of the input $\kappa$ and the ${\cal E}$ and ${\cal S}$
statistics. The results correspond to a
$100^{\circ} \times 100^{\circ}$ simulation observed with an ideal CMB experiment
with negligible detector noise and infinite angular resolution.} 
\label{crosscorr1}
\end{figure*}

The cross correlations with the input $\kappa$ give
\begin{eqnarray}
C_l^{\cal S \kappa} &=& 2C^{\kappa \kappa}_{l} \nonumber \\
C_l^{\cal E \kappa} &=& 2C^{\kappa \kappa}_{l} \nonumber \\
C_l^{\cal B \kappa} &=& 0.
\label{cross}
\end{eqnarray}

Figure \ref{crosscorr1} shows the cross correlations in equation
(\ref{cross}) for a
$100^o\times 100^o$ field where the CMB is measured with negligible
detector noise and infinite angular resolution (the results are almost the same
for Planck specifications). The agreement is remarkable, proving that
there is enough signal in the $\kappa$ maps recovered using our
technique to get useful information about the convergence power
spectra by cross correlating them with other maps. This is not surprising, 
since there are about 5$\times 10^5$ independent patches of size $\xi^2=(0.15^{\circ})^2$
and we need about $(N_l^{\cal EE}/4C_l^{\kappa \kappa})^2 \sim 100-400$ of them to obtain 
signal to noise of unity.

In reality most tracers of the underlying mass will not be perfect
because they correlate only partially with $\kappa$. 
In particular we assume that we have a map $Y$ that correlates with
$\kappa$ and has a cross correlation $C^{\kappa Y}_{l}$. 
Then the cross correlations between the 
${\cal W}={\cal  S, E}$ maps and the $Y$ map give,
\begin{eqnarray}
C^{{\cal W} Y}_l&=& 
 2C^{\kappa Y}_{l} W(l) 
\label{cross2}
\end{eqnarray}
where $W(l)$ is the window defined in equation (\ref{window}).
The correlation of the $Y$ map with $\cal B$ vanishes,
$C^{{\cal B} Y}_l= 0$.
The covariance matrix for the two correlations is,
\begin{eqnarray}
{\rm Cov}(C^{\cal W^{\prime} Y}_l C^{\cal W
Y}_l)&=&(C_l^{YY}+N_l^{YY})(4C_l^{\kappa \kappa} W^2(l) + N_l^{\cal W^{\prime}
W})+ 4C^{\kappa Y}_{l}C^{\kappa Y}_{l} W^2(l), 
\end{eqnarray}
where the first term is usually dominant. Both the decorrelation and the 
noise reduce the overall signal to noise of the cross correlation.
For example, in the case of cross correlation with the CMB only the largest modes are 
strongly correlated because the time dependent gravitational potential does 
not contribute to the CMB anisotropies 
on small scales. The overall signal to noise is significantly reduced
compared to the idealized example discussed here, 
but is nevertheless detectable with future satellite missions for 
reasonable low density models \cite{isw}.

\section{Conclusions}

We have developed a method to reconstruct projected mass density from observed 
cosmic microwave background maps. The method consists of taking derivatives
of temperature field and averaging their products. Particular combinations 
on average correspond to shear and convergence of gravitational lensing, 
which can be expressed as a line of sight integral over the density field.

We have presented three possible applications of the method. First we 
applied the method to simulated clusters showing 
that it can be successfully used
to reconstruct their projected  mass distribution if there is sufficient 
small scale power. For this to be successful we require small scale 
power beyond the one given by primary anisotropies, which could be provided 
by secondary processes such as Ostriker-Vishniac effect or primeval galaxies.
The expected signal from larger structures, such as 
filaments and superclusters, is smaller and therefore more difficult 
to directly observe.

A second and more promising application is to average the reconstructed map
to extract the power spectrum of fluctuations.
In simulations
the method successfully reconstructs the input power spectrum both
in the large scale limit (where we are averaging over many independent 
patches of CMB to reduce the noise) and in the small scale limit (where
lensing generates additional small scale power). Future satellite missions
should be able to measure this signal at least on large scales
\cite{letterps} and provide additional constraints on the cosmological parameters.  
Interferometers may be able to determine the power spectrum on smaller scales
as well. This method is complementary and in many cases more robust
than the more traditional methods of 
determining the power spectrum. In comparison to the power spectrum from galaxy 
surveys the main advantage is that there is no assumption on how light traces mass, 
which is still poorly understood at present.
In comparison with the weak lensing surveys planned for the 
future the advantage is that the redshift distribution of the source is well known
($z \sim 1100)$ and much higher than for any galaxy survey. It also does not suffer from
intrinsic correlations between the galaxies, which may mimic the weak lensing signal.
In comparison with other direct tracers of dark matter such as velocity flows and 
Ly-$\alpha$ forest the method presented here recovers the power spectrum over 
a larger range in scale and is
less sensitive to basic assumptions underlying the method such as the error distribution of 
galaxy distances or assumptions of the IGM at high redshift.
In addition, the method presented here also gives power spectrum information on much
larger scales than reachable by other methods. These are likely to be linear and
may allow us to
deconvolve the power spectrum to obtain the 3-d power spectrum using the methods 
developed in \cite{seljak98}.

A third promising application of the reconstruction is its cross correlation with other 
maps that trace
large scale structure. Some of these are X-ray background, galaxy surveys and 
CMB itself (both the primary anisotropies and SZ contribution). 
This again allows us to average over many independent patches to 
reach a positive detection and may give even higher statistical significance 
than the power spectrum if the two maps are well correlated. In this
case we can 
learn not only about the dark matter clustering, but also about its relation 
to X-ray and optical light or about a time dependent gravitational potential.
Overall, CMB fluctuations may hide a whole new information 
treasure in its pattern and its extraction would provide important information
on the universe we live in.

\smallskip
U.S. and M.Z. would like to thank Observatoire de Strasbourg and MPA, 
Garching, respectively, for 
hospitality during the visits.  
M.Z. is supported by NASA through Hubble Fellowship grant
HF-01116.01-98A from STScI,
operated by AURA, Inc. under NASA contract NAS5-26555.

\end{document}